\documentclass[twocolumn,preprintnumbers,amsmath,amssymb, superscriptaddress,nofootinbib]{revtex4}

\usepackage{graphicx}
\usepackage{dcolumn}
\usepackage{bm}
\usepackage{bbm}
\usepackage{doi,natbib}
\usepackage{url, xcolor}

\usepackage{amsmath}
\usepackage{amsthm}
\usepackage{amssymb}
\usepackage{array}
\usepackage[latin10]{inputenc}

\usepackage{amsfonts}
\usepackage{amssymb}
\usepackage[english]{babel}

\usepackage[font={small,it}]{caption}

\newcommand{\avg}[1]{\langle #1\rangle}
\newcommand{\bra}[1]{\langle #1|}
\newcommand{\ket}[1]{|#1\rangle}

\newcommand{\ketbra}[2]{|#1\rangle\!\langle#2|}
\newcommand{\id}{\mathbbm{1}}

\DeclareMathOperator{\tr}{Tr}

\newcommand\rev{\mathrm{rev}}
\newcommand\fwd{\mathrm{fwd}}

\newcommand\mmod{\operatorname{mod}}

\newcommand{\sneg}{\hspace{-0.1em}}

\let\oldmarginpar\marginpar
\renewcommand\marginpar[1]{\-\oldmarginpar[\raggedleft\marginparsize #1]%
{\raggedright\marginparsize #1}}

\begin{document}

\setlength{\tabcolsep}{1ex}

\title{Entropic equality for worst-case work at any protocol speed}

\author{Oscar C. O. Dahlsten}
\affiliation{Blackett Laboratory,
Imperial College London SW7 2AZ, UK}
\affiliation{Clarendon Laboratory,
University of Oxford, Parks Road, Oxford OX1 3PU, UK}
\affiliation{LIMS, 35A South St, Mayfair, London W1K 2XF, UK}
\author{Mahn-Soo Choi}
 \affiliation{Department of Physics, Korea University, Seoul 136-701, South Korea}
 \author{Daniel Braun}
 \affiliation{ITP, Universit\"at T\"ubingen, Auf der Morgenstelle 14, 72076 T\"ubingen, Germany}
\author{Andrew~J.~P.~Garner}
\affiliation{Clarendon Laboratory,
University of Oxford, Parks Road, Oxford OX13PU, United Kingdom}
\affiliation{Center for Quantum Technologies, National University of Singapore, Republic of Singapore}

\author{Nicole~Yunger~Halpern}
\affiliation{Institute for Quantum Information and Matter, Caltech, Pasadena, CA 91125, USA}

\author{Vlatko Vedral}
\affiliation{Clarendon Laboratory,
University of Oxford, Parks Road, Oxford OX13PU, United Kingdom}
\affiliation{Center for Quantum Technologies, National University of Singapore, Republic of Singapore}
\date{\today}

\begin{abstract}
We derive an equality for non-equilibrium statistical mechanics in finite-dimensional quantum
systems. The equality concerns the worst-case work output of a time-dependent Hamiltonian protocol in the presence of a Markovian heat bath. It has has the form``worst-case work = penalty -
optimum'' The equality holds for all rates of changing the Hamiltonian and can be used to derive
the optimum by setting the penalty to 0. The optimum term contains the max entropy of the
initial state, rather than the von Neumann entropy, thus recovering recent results from single-shot
statistical mechanics. Energy coherences can arise during the protocol but are assumed not to be
present initially. We apply the equality to an electron box.
\end{abstract}

\maketitle
\noindent{\bf {\em  General introduction---}}Average values of quantities are not always typical values. In non-equilibrium nano and quantum systems this is often the case, with, for example, the work output of a protocol having a significant probability of deviating from the average. Hence, in these important systems, statements about averages have limited use when it comes to predicting what will happen in any given trial; the fluctuations need to be discussed explicitly. Two key relations concerning fluctuations in work, Crooks' Theorem~\cite{Crooks99} and Jarzynski's Equality~\cite{Jarzynski97}, have been studied extensively theoretically and experimentally.  These theorems hold for any speed of changing the Hamiltonian, the thermalisation can be partial or negligible during the protocol. Amongst other things the theorems can be used to determine free energies of equilibrium states from non-equilibrium experiments. 

A recently developed alternative approach is \emph{single-shot statistical mechanics}~\cite{DahlstenRRV11,DelRioARDV11,HorodeckiO13,Aberg13,
EgloffDRV12,FaistDOR12, Dahlsten13,BrandaoHNOW13,YungerHalpernR14,BrowneGDV14}, inspired by single-shot information theory~\cite{Renner05, RennerW04}. The focus is on statements that are guaranteed to be true in every trial, rather than on average behaviors. For example, one can ask whether a process's work output is guaranteed to exceed some threshold value (such as an activation energy), or whether a process's work cost is guaranteed not to exceed some threshold value (beyond which the system may break from dissipating heat). These statements concern the \emph{worst-case work} of a process. A key realization is that the optimal worst-case work is determined not by the von Neumann/Shannon entropy of the initial state, but rather the max entropy, which is the logarithm of the number of non-zero eigenvalues of the density matrix. Thus, which entropy one should use in statements about optimal work depends on which property of the work probability distribution one is interested in.  

\begin{figure}
\centering
\includegraphics[width=0.8\linewidth]{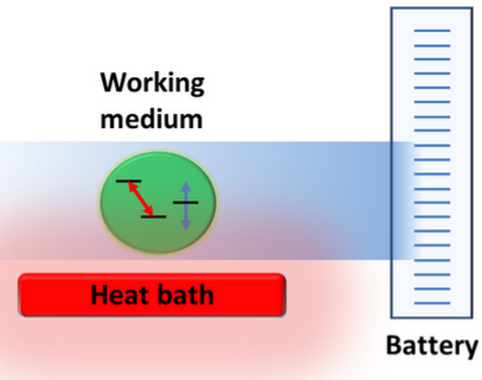}
\caption{Setup: a working-medium,
a battery 
from which work is taken or given to, and a single heat bath. The
battery system alters the Hamiltonian of the working-medium, depicted
with the blue arrow shifting an energy level. The heat bath 
induces jumps between the system energy levels, depicted by the red arrow.}
\label{fig:setup}
\end{figure}

Single-shot statistical mechanics began with almost no \emph{a priori} relation to fluctuation theorems. Promising links were made in~\cite{Aberg13,YungerHalpernGDV14}.  In \cite{YungerHalpernGDV14} it was shown that Crooks' Theorem can be used to make statements about worst-case work. This created the beginnings of a bridge between Crooks' theorem and results in single-shot papers.  We here complete the bridge, showing that key expressions concerning optimal worst-case work from~\cite{DahlstenRRV11,HorodeckiO13,Aberg13} follow from Crooks' Theorem plus some extra thought.  To our knowledge this is a new and unexpected result. We moreover generalize them by giving an equality for the worst-case work. The equality governs 
time-varying Hamiltonian protocols, including fast ones, assuming Markovian heat baths and a weak restriction on the strength of any initial quench.
The initial state is taken to be diagonal in the energy eigenbasis, though not necessarily thermal, and energy coherences can arise during the protocol. 
Figure~\ref{fig:setup} illustrates the set-up.
The equality has the form `worst-case work$\; =$ penalty - optimum,' with the penalty term guaranteed to be non-negative such that the optimum can be derived by setting the penalty to zero. We believe this concrete link to fluctuation theorems will give significant impetus to single-shot statistical mechanics, allowing it to harness results from the highly developed fluctuation theorem approach.
As a demonstration we apply the result to an electron box experiment described in the fluctuation theorem formalism~\cite{Koski14a, Koski13a, Saira12a}.  We begin by defining the general set-up.

\noindent{\bf {\em Trajectory model---}}The physical scenario is depicted in Fig.~\ref{fig:setup}.
For concreteness, we shall use the following explicit model (and later discuss other possible models). A protocol will be a sequence of elementary changes: (i) changes of the Hamiltonian and (ii) thermalizations. We shall initially assume there is a finite number of such steps (but later show that the continuum limit is well-defined and corresponds to a master equation, at least in the discrete-classical case). The Hamiltonian is parameterized by $\lambda_m$, with $m$ an integer that labels the step.

{\bf 1. Hamiltonian changes} map $\lambda_m$ to $\lambda_{m+1}$. 
We follow~\cite{QuanD08} in supposing there is an energy measurement in the instantaneous energy eigenbasis at the beginning and end of each Hamiltonian-changing step. In a given realization the system then evolves from $\ket{i_m,\lambda_m}$ to $\ket{i'_m,\lambda_{m+1}}$, where $i_m$ labels the energy eigenstate. This costs work given by the energy difference: $w_m  =  E(\ket{i_m,\lambda_{m+1}} )  -   E(\ket{i'_m,\lambda_{m}})$. An important special case is $i_m=i'_m$, which arises in the quasi-static (quantum adiabatic) limit, as well as if the energy eigenbasis is constant and only the energy eigenvalues change; this can be termed the discrete-classical case. 

{\bf 2. Thermalizations} map $i'_m$ to $i_{m+1}$, cost no work, and preserve the Hamiltonian: $\ket{i'_m,\lambda_{m+1}}\rightarrow \ket{i_{m+1},\lambda_{m+1}}$. For notational simplicity let us label this as 
 $\ket{i}\rightarrow \ket{j}$ with energy $E_i \rightarrow E_j$.
We do not assume that the system thermalizes fully but that the hopping probabilities respect thermal detailed balance:
$\frac{p\left(\ket{i}\!\rightarrow\!\ket{j}\right)}
{p\left(\ket{j}\!\rightarrow\!\ket{i}\!\right)}
\!\!=\!\sneg
e^{   -\!\beta( E_j-E_i)}.
$ The energy change $E_j-E_i$ from such a step is called heat, $Q_m$.

A \emph{trajectory} is the time-sequence of energy eigenstates occupied:
$\ket{i_0,\lambda_0}\rightarrow \ket{i'_0,\lambda_{1}} 
\rightarrow \ket{i_1,\lambda_1}
\rightarrow  \ldots
\rightarrow \ket{i_{f-1},\lambda_{f-1}}
\rightarrow \ket{i'_{f-1},\lambda_{f}}
\rightarrow  \ket{i_{f},\lambda_{f}}.$

The probability of a given trajectory is accordingly, assuming a Markovian heat bath,
\begin{eqnarray}
\label{eq:trajprob}
p(traj)&=&p(\ket{i_0,\lambda_0})  \prod_{m=0}^f\! \!p(\ket{i_m,\lambda_m}\!\rightarrow \!\ket{i'_m,\lambda_{m+1}})   \nonumber \\  
&\times & p(\ket{i'_m,\lambda_{m+1}}\!\rightarrow \!\ket{i_{m+1},\lambda_{m+1}}). 
\end{eqnarray}
A trajectory's {\em inverse} is the reverse of the sequence. The
inverse corresponds, in the discrete-classical case, to the
Hamiltonian changes running in reverse, from $\lambda_f$ to
$\lambda_0$, and to the same thermalizations as in the forward
protocol, with the sequence exactly inverted. This process is termed
the reverse process. Beyond the discrete-classical case, 
the Hamiltonian changes corresponding to the reverse process are
defined such that $p(\ket{i'_m,\lambda_{m+1}}\rightarrow
\ket{i_m,\lambda_m})  
=   p(  \ket{i_m,\lambda_m} \rightarrow \ket{i'_m,\lambda_{m+1}}
)$. Our results will hold under that condition. There are at least two
ways of satisfying that condition: (i) Simply let the unitary of the
corresponding elementary step in the reverse process be $U^{-1}$,
where $U$ is that of the forward 
process, (ii) apply a suitable `time-reversal' operator $\Theta$ to all states and operators involved, as in~\cite{QuanD08}. The reverse trajectory is then the reverse sequence of the time-reversed energy eigenstates: $\Theta \ket{i_{f},\lambda_{f}}... \Theta \ket{i_{0},\lambda_{0}}$, with the condition $p(\ket{i'_m,\lambda_{m+1}}\rightarrow \ket{i_m,\lambda_m})
=   p(  \Theta \ket{i_m,\lambda_m} \rightarrow \Theta \ket{i'_m,\lambda_{m+1}}  )$ being satisfied, as time reversal implies taking the complex conjugate of the states, in a preferred basis, and the transpose of the time-evolution in the same basis: $U\rightarrow U^T$. The condition is thus satisfied as $\bra{b} U \ket{a}=(\bra{a}U^\dagger \ket{b})^*=\bra{a}^*U^T\ket{b}^*$.

A given trajectory has some work cost $w=\sum_m w_m$, in line with the
definition of the Hamiltonian-changing steps. The inverse trajectory
has work cost $-w$. A given protocol on a given initial state induces
a probability distribution over trajectories, with an associated
probability distribution over work $p(w)$. The forward  
and reverse protocols give rise to $p_{\fwd}(w)$ and $p_{\rev}(-w)$ respectively.

If the initial density matrices of the forwards and reverse processes are both thermal,  
$\exp{\bm (}- \beta H(\lambda_0) {\bm )}/Z_0$ and $\exp{\bm (}- \beta H(\lambda_f) {\bm )}/Z_f$,  Crooks' Theorem holds~\cite{QuanD08}:
\begin{equation}
\label{eq:Crooks}
\frac{p_{\fwd}(w)}{p_{\rev}(-w)}=\frac{Z_f}{Z_0} \exp (\beta w).
\end{equation}
 (To derive it take the ratio of Eq.~\ref{eq:trajprob} and the corresponding reverse trajectory expression. Apply thermal detailed balance and the equality of reverse hopping probabilities for the Hamiltonian-changing steps. Sum over trajectories with the same $w$, and note that the reverse of a trajectory has the same work up to a minus sign~\cite{QuanD08}.)

\noindent{\bf {\em  Worst-case work---}}The central object of our interest is the \emph{worst-case work}
\[w^0 :=  \max  \{  w : p(w)> 0  \}, \]
also known as the \emph{guaranteed work}~\cite{EgloffDRV12}. In physical situations this will have a finite upper bound as no battery has infinite energy.
The worst-case value may be realized by a very unlikely trajectory. It 
is then natural to consider the worst-case work of a subset of trajectories $\mathcal{T}$:
\[w^0_\mathcal{T} :=  \max  \{  w : p(w)> 0 \text{ and traj} \in \mathcal{T} \}. \]

\noindent{\bf {\em One-shot relative entropies---}}The standard relative entropy is
$D( \rho || \sigma )  :=  -  {\rm Tr} (\rho [ \log \rho  -  \log \sigma ] )$~\cite{NielsenC10}, wherein $\log$ denotes (in this paper) the natural logarithm, or $\ln$. This $D$ belongs to a class of relative entropies known as the \emph{R\'{e}nyi relative entropies}, parameterized by $\alpha \in \mathbb{R}$. We shall use two other members of that family: the (classical version of the) $\infty$-relative entropy
$D_\infty ( P || Q )  :=  \sup_x \log     ( \frac{p_x}{q_x})$ and the $0$-relative entropy
$D_0( \rho || \sigma )  :=  - \log{\rm Tr} ( \pi_\rho \sigma)$, wherein $\pi_\rho$ 
projects onto the support of $\rho$~\cite{Datta09}. These are called \emph{one-shot relative entropies}, as they arise naturally in one-shot (or single-shot) information theory~\cite{Renner05, RennerW04,Datta09}.

\noindent{\bf {\em Worst-case work from Crooks---}} It was shown in~\cite{YungerHalpernGDV14} that one can recover an expression for the worst-case work $w^0$ from Crooks' Theorem (Eq.~\ref{eq:Crooks}.  Consider the equality of Crooks' Theorem (for
values of $w$ such that $p_{\fwd}(w)>0$) and select the value for $w$ which
maximizes the LHS (and thus the RHS)~\cite{YungerHalpernGDV14}: 
$$\max \frac{p_{\fwd}(w)}{p_{\rev}(-w)}=\max \frac{Z_f}{Z} e^{ \beta w}.$$
The RHS is monotonic in $w$, so maximizing the RHS over the support of $p_{\fwd}(w)$ leads to the maximum $w$-value $w^0$. Taking the logarithm and recalling the $D_\infty$ definition yields: 
\begin{equation}
\label{eq:premain}
\beta w^0 = D_{\infty} (p_{\rm fwd}(w) || p_{\rm rev}(-w))-\log\left(Z_f/Z\right).  
\end{equation}
Note that this derivation assumes Crook's theorem which does not in general hold for athermal initial states. 

\noindent{\bf {\em Equality for worst-case work---}} Consider an initial state $\rho_0$, and a protocol of thermalizations and Hamiltonian changes with initial and final Hamiltonians $H(\lambda_0)$ and $H(\lambda_f)$ respectively.  This induces a work probability distribution $p(w)$ and an associated $w^0$. We shall derive an equality of the form $w^0\; =$ penalty - optimum.

We consider initial states of form $\rho_0=\sum_i p_i \ket{i_0,\lambda_0}\bra{i_0,\lambda_0}$, i.e., diagonal in the energy eigenbasis though not necessarily thermal (energy coherence may still arise during the protocol). We take $p_i \neq 0$. This is because we wish to avoid divergences from dividing by $p_i$. (See~\cite{MurashitaFM14} for an alternative approach to this divergence problem.)

To apply Crooks' Theorem (Eq.~\ref{eq:Crooks}) here, even though the initial state is not assumed to be thermal, our approach is as follows. For example, if one has a degenerate two-level system, the thermal state is $\gamma=1/2 \ket{0}\bra{0}+1/2 \ket{1}\bra{1}$. If one instead had $\rho_0=2/3 \ket{0}\bra{0}+1/3 \ket{1}\bra{1}$, the worst-case work would be the same as for $\gamma$. This follows because the set of trajectories with nonzero probability is the same in both cases, as can be seen from Eq.~\ref{eq:trajprob} which gives the probability of a trajectory. Given a $\rho_0$, we will find a corresponding thermal state with the same worst-case work and apply Crooks' Theorem to that. 

An important practical consideration which makes this more subtle is that some $p_i$ may be negligible and even arbitrarily close to 0. It is natural to exclude trajectories starting in those states when calculating the worst-case work. 
We therefore divide the initial energy eigenstates into two sets.
One set is the one of interest: $\mathcal{E}_\text{IN}$.
The set of the other eigenstates, we call $\mathcal{E}_\text{OUT}$,
corresponding to those we shall exclude when calculating the worst-case work. 
The probability of being in $\mathcal{E}_\text{OUT}$ is given by  
$$p(\text{OUT})=\sum_{\ket{i_0,\lambda_0}\in \mathcal{E}_\text{OUT}} \tr(\ket{i_0,\lambda_0}\bra{i_0,\lambda_0}\rho_0) .$$
We define $\mathcal{T}_\text{IN}$ as the set of possible ($p>0$)
trajectories beginning in $\mathcal{E}_\text{IN}$.
Similarly, we define
$\mathcal{T}_\text{OUT}$ as the set of possible trajectories beginning in
$\mathcal{E}_\text{OUT}$. 
Recall that each trajectory corresponds to some work value. 
We call the worst-case work of $\mathcal{T}_\text{IN}$, $w^0_\text{IN}$.
This cannot be worse than the worst-case over all trajectories: $w^0_\text{IN}\leq w^0$.

Let us design an associated thermal state that yields the same worst-case work as $\rho_0$: $w^0_\text{IN}$.
Later, we show that this is indeed the case, under an additional mild
assumption. 
The associated thermal state has the same Hamiltonian as the system apart from the OUT levels. We define the Hamiltonian as $\widetilde{H}:=\sum_\text{IN}E_i\ket{i}\bra{i}+\sum_\text{OUT}\widetilde{E_i}\ket{i}\bra{i}$, changing the energies of the states in $\mathcal{E}_\text{OUT}$ to new ones, $\widetilde{E}_i$, such that $p_i=\exp(-\beta \widetilde{E}_i)/\widetilde{Z}$, and leaving the other energy levels the same.
The thermal state associated with that Hamiltonian is then 
\begin{equation*}
\widetilde{\gamma}\!=\!\!\!\!\sum_{\ket{i_0,\lambda_0}\in \mathcal{E}_\text{IN}}\! \!\!\frac{e^{-\beta E_i}}{\widetilde{Z}}\ket{i_0,\lambda_0}\bra{i_0,\lambda_0}+\!\!\!\!\!\!\sum_{\ket{i_0,\lambda_0}\in \mathcal{E}_\text{OUT}} \!\!\!p_i\ket{i_0,\lambda_0}\bra{i_0,\lambda_0},
\end{equation*}
 The definition implies that
\begin{equation}
\widetilde{Z}=\frac{\sum_{\ket{i_0,\lambda_0}\in \mathcal{E}_\text{IN}} e^{-\beta{E_i}}}{1-p(OUT)}.
\end{equation}
This partition function differs from that of the actual Hamiltonian
$H(\lambda_0)$. 

In this scenario with $\widetilde{\gamma}$ as the initial state and
the $\mathcal{E}_\text{OUT}$ levels lifted, the protocol is the same as
in the actual scenario, except that initially the energies of the
states in $\mathcal{E}_\text{OUT}$ 
 are lowered down to the levels of the actual Hamiltonian of interest. The worst-case work of this scenario is called $\widetilde{w}^0$. Under a mild additional restriction on protocols considered, roughly speaking that the worst-case work is bounded from below---as is the case for physically realisable protocols (see Methods), we then have 
\begin{equation}
\label{Eq:Worstcaseequiv}
\widetilde{w}^0=w^{0}_\text{IN}.
\end{equation}
 To get $\widetilde{w}^0$ from Crooks' Theorem (Eq.~\ref{eq:Crooks}) we
shall make use of Eq.~\ref{eq:premain} from~\cite{YungerHalpernGDV14}. This applies in the scenario with $\widetilde{\gamma}$ as the initial state, as Crook's theorem holds in that scenario (see discussion around Eq.~\ref{eq:Crooks}), and thus
\begin{equation}
\label{eq:premaintilde}
\beta \widetilde{w}^0 = D_{\infty} (\widetilde{p}_{\rm fwd}(w) || \widetilde{p}_{\rm rev}(-w))-\log\left(Z_f/\widetilde{Z}\right).  
\end{equation}
\noindent{\bf {\em Main result}---} Combining Eq.~\ref{eq:premaintilde} and Eq.~\ref{Eq:Worstcaseequiv} we have
\begin{equation}
\label{eq:mainbodymain}
\beta w^0_\text{IN} = D_{\infty} (\widetilde{p}_{\rm fwd}(w) || \widetilde{p}_{\rm rev}(-w))-\log\left(Z_f/\widetilde{Z}\right).  
\end{equation}
Thus the worst case work of the trajectories of interest
$w^0_\text{IN}$ is this equal to (kT times) a relative entropy minus
(the logarithm) of the ratio of two partition functions,
one of which encodes information about how many of the initial energy eigenstates have negligible occupation probability.

\noindent{\bf{\em  Discussion---}}
Note that Equation~\ref{eq:mainbodymain} has the
form $$\beta w^0=\text{ penalty - optimum}.$$ The penalty is essentially given by the
difference between the forward and reverse distributions, quantified
by $D_{\infty}$. The optimum one can hope for, with a given initial
state and given initial and final Hamiltonian, is to set the penalty
to 0 (as relative entropies are non-negative), which leaves
$-\log\left(Z_f/\widetilde{Z}\right)$. This term is 
more
negative the smaller the support of $\rho$ is and the lower the final
energies are relative to the initial ones.  This optimum is achieved by 
a protocol in~\cite{EgloffDRV12}.

We now consider the optimum term in two important special cases where the single-shot entropy of the initial state emerges. To simplify the considerations we here set $p(\mathrm{OUT})\approx 0$, although our full expressions do not assume that to be the case. The `epsilonics' are dealt with explicitly in the Methods. (i)  Consider firstly the special case of  $H(\lambda_0)=H(\lambda_f)$ which has been studied in the single-shot statistical mechanics literature. Then  
$\log  \left(Z_f/\widetilde{Z}\right)=-\log \sum_{i\in IN }
e^{-\beta E_i}/Z_f$. This can be rewritten, informally, using the definition of $D_0$ in the technical introduction and noting that the sum is only over IN levels, as $D_0(\rho_0^{(IN)}||\gamma)$, where $\rho_0^{(IN)}$ is $\rho_0$ with the probability tail in OUT cut off---this is made general and precise in the discussion on smooth relative entropies in  Appendix \ref{app.smooth}. Thus in this case the equality of Eq.\ref{eq:mainbodymain} has the form
$$\beta w^0=D_{\infty}-D_0.$$ 
 (Recall $D_{\infty}$ concerns work distributions and $D_0$ states.)
 (ii) Further restricting the Hamiltonian such that 
$H(\lambda_0)=H(\lambda_f)=0$, we have  $D_0(\rho_0||\gamma)=\log
d-S_{\max}(\rho_0)$ (noting $\gamma=\id/d$ and recalling $S_{\max}(\rho):=S_0(\rho):=\log(\text{rank}(\rho))$. This recovers the known
results from~\cite{DahlstenRRV11,HorodeckiO13,Aberg13} that these are
optimal in the respective cases.  The message is that it is the max entropy $S_{\max}$ which
determines the optimal worst-case work, rather than the von Neumann entropy. If one defines thermodynamic entropy in terms
of optimally extractable {\em worst-case} work, it is the max entropy which should be used.

\begin{figure}
\centering
\includegraphics[width=5cm]{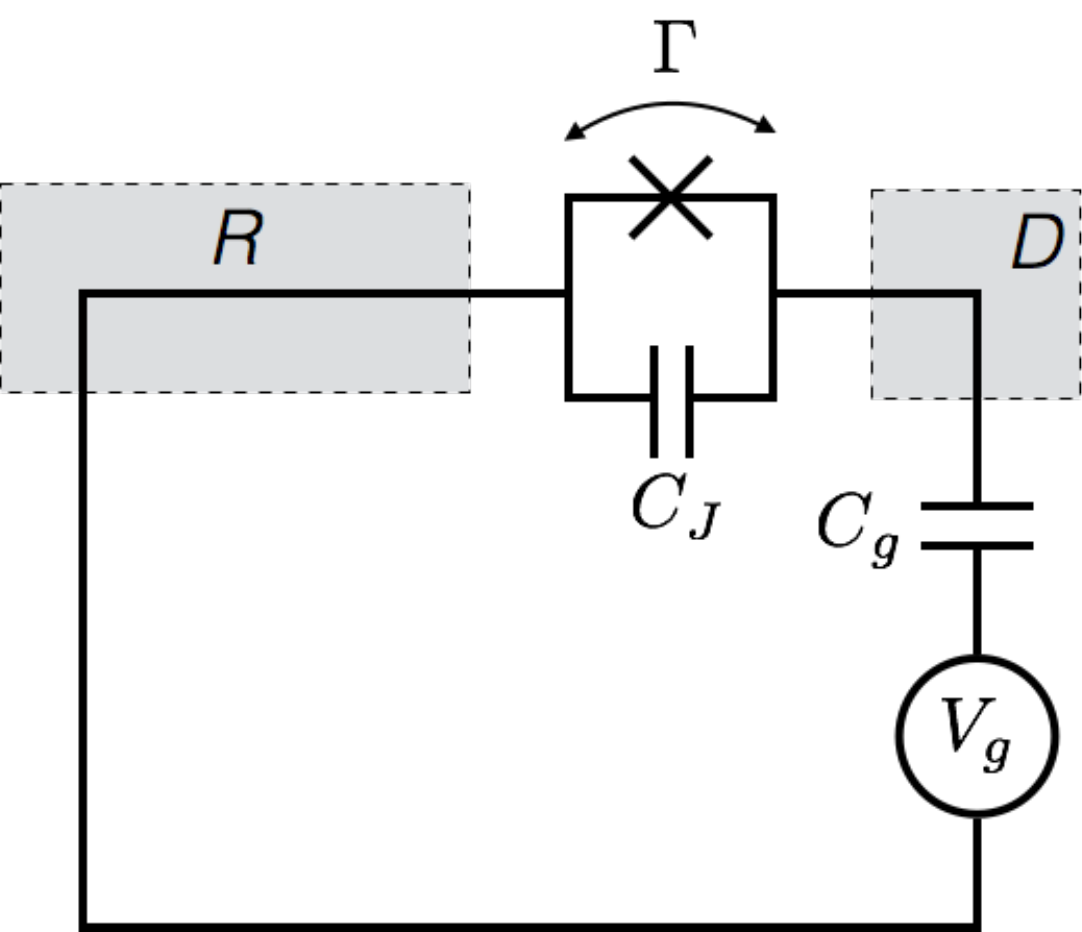}
\caption{ An ``electron box'' (D) coupled to a metallic electrode (R) via tunnelling and the capacitor with capacitance $C_J$, and to the gate electrode via the capacitor with $C_g$. The gate voltage $V_g$ controls the number of excess electrons in the electron box, $N$. At low temperatures N is restricted to two possible values associated with $\ket{0}$ and $\ket{1}$, with relative energy $H\propto -C_gV_g\ket{1}\bra{1}$. The electrode $R$ plays the role of a heat bath, with tunnelling in/out of D corresponding to thermal excitation/relaxation. Experimentally the work and heat can be measured by probing the charge on $D$ in real time~\cite{Saira12a,Koski13a,Koski14a}.
}
\label{fig:electronbox}
\end{figure}

To make the connection to physics clear, we apply the results to a
recent realization of a Szilard engine with an electron
box~\cite{Koski14a, Koski13a, Saira12a}. A great advantage 
of using this trajectories model from the fluctuation theorem approach is that it allows the application of single-shot results to such experiments. We described the set-up in Fig.~\ref{fig:electronbox} and in the Methods section, we analyze what controls the penalty term $D_{\infty}$ in this scenario. We also describe in the Methods how the penalty term, up to vanishing probabilistic error, goes to 0 in the isothermal quasistatic limit. 

As described in the {\em trajectories} section, these results also
apply if the evolution includes unitaries that create energy
coherences, including sudden changes of the energy eigenbasis such that a state that 
was orginally an energy eigenstate is now classified as a superposition of energy eigenstates. 
Such coherences are normally viewed as associated with entropy production and extra work costs. 
We note an interesting counter-example. Working within this trajectory model, 
suppose $H(\lambda_0)=0$;
$\rho_0=1/3\ket{0}\bra{0}+2/3\ket{1}\bra{1}$, and
$H(\lambda_f)=\delta E\ket{i_f}\bra{i_f}$. 
 If the energy eigenstates stay the same throughout such that
 $\ket{i_f}=\ket{1}$, the worst-case work is $\delta E$, and it has
 probability $2/3$  (even if the shift is done quickly). If instead
 the Hamiltonian eigenstates change such that $\ket{0}\rightarrow
 \ket{+}$, and $\ket{1}\rightarrow \ket{i_f}= \ket{-}$
then the worst-case work is still $\delta E$, corresponding to outcome $\ket{-}$ of the final energy measurement. However the probability of this can be as low as 1/2 (if $H$ is changed suddenly $p(\ket{-})=\tr(\rho_0 \ket{-}\bra{-})=1/2$). This shows that the probability of the worst case can actually be improved (lowered) by coherence due to a sudden change of the Hamiltonian.
This improvement comes at the cost of randomising the work distribution. 

The derivation of the main result relies very little on the specifics of the trajectory model.
It would e.g. also go through with the quantum-jump type model in~\cite{HekkingP13, inprep}.
That model uses no intermediate projective measurements on the system but rather on the heat bath, as is natural in quantum optics.

\noindent{\bf {\em Summary and outlook}---}We showed that in any protocol with a time-varying Hamiltonian and thermalizations, the worst-case work takes the form of ``penalty - optimum". The model we used could be generalized in various ways, including non-Markovian baths and baths that decohere in other bases than the energy basis. It is also important to find more bounds for the penalty term in terms of controllable parameters. Finally we note that the results of~\cite{Perry15} suggest that the bounds obtained here also apply to what is known as thermal operations in the context of resource theories---another interesting question.

\noindent{\bf {\em Note added}---}Similar results were obtained independently by Salek and Wiesner, using a different set-up and different starting assumptions, in: {\em Fluctuations in Single-Shot $\epsilon$-deterministic Work Extractions} ~\cite{SalekW16}.

\noindent{\bf {\em Acknowledgements}---}We are grateful to comments on a draft by Dario Egloff. We acknowledge funding from the EPSRC (UK), the Templeton Foundation, the Leverhulme Trust, the Oxford Martin School, the National Research Foundation (Singapore), the EU collaborative project TherMiQ (Grant agreement No.~618074), the Ministry of Education (Singapore), NSF grant PHY-0803371, an IQIM Fellowship, and a Virginia Gilloon Fellowship, a  BK21 Plus Project from the Ministry of Education of Korea, and  the Gordon and Betty Moore Foundation.


\onecolumngrid
\appendix

\section{Properties of $\widetilde{\gamma}$ and associated protocol}
 As Crooks' theorem, which we use, is for initial thermal states, we designed a thermal state $\widetilde{\gamma}$, with the aim of 
creating a set-up which yields the same worst-case work as for the actual initial state, which is not necessarily thermal. 
For a given initial state $\rho_0=\sum p_i\ket{i}\bra{i}$ and initial energy eigenvalues $E_i$, the associated thermal state is defined as
 $\widetilde{\gamma}=\sum_{i}
 e^{-\beta{\widetilde{E_i}}}/\widetilde{Z}\ket{i}\bra{i}$, where
 $\widetilde{E_i}=E_i$ for $\ket{i}\in \mathcal{E}_\text{IN}$, but for
 $\ket{i}\in \mathcal{E}_\text{OUT}$, $\widetilde{E_i}$ is chosen such that
 $e^{-\beta{\widetilde{E_i}}}/\widetilde{Z}=p_i $. Physically,  this implies
 replacing the energy levels with small occupation probability $p_i$ by
 much higher energy levels such that their thermal occupation
 probability is as small as $p_i$. The Hamiltonian associated 
 with $\widetilde{\gamma}$ is accordingly
 $\widetilde{H}:=\sum_\text{IN}E_i\ket{i}\bra{i}+\sum_\text{OUT}\widetilde{E_i}\ket{i}\bra{i}$. 
 The normalising factor is $\widetilde{Z}=\sum_{\ket{i}} e^{-\beta{\widetilde{E_i}}}$. These definitions imply that
\begin{equation}
\label{eq:tildez}
\widetilde{Z}=\frac{\sum_{\ket{i}\in \mathcal{E}_\text{IN}} e^{-\beta{E_i}}}{1-p(OUT)}.
\end{equation}

Apart from the given actual protocol, we then also design a $\sim$-protocol
such that it gives the same worst-case work in the case of
$\widetilde{\gamma}$ as the initial state.  
We define the $\sim$-protocol as beginning with  $\widetilde{H}$, then
lowering the OUT levels back to $E_i$, i.e., setting
$\widetilde{H}\rightarrow H$. After that it is the same as the actual
protocol. We call the $\sim$-protocol applied to $\widetilde{\gamma}$ ``the
$\sim$-scenario.''

In the $\sim$-scenario we similarly have $\widetilde{\mathcal{T}}_\text{IN}$ and $\widetilde{\mathcal{T}}_\text{OUT}$, and $\widetilde{w}^0_\text{IN}$. The following holds:
\begin{equation}
\label{eq:worstcaseequiv}
\widetilde{w}^0_\text{IN}=w^0_\text{IN},
\end{equation}
i.e., the worst-case work is the same in the $\sim$-scenario as in the
actual scenario, for the $\mathcal{T}_\text{IN}$ subset of trajectories. 
This is because the protocol is defined above such that the added
initial step in the $\sim$-scenario only involves shifting the OUT
levels (without any thermal hopping). The set of possible work values
is 
the same in $\mathcal{T}_\text{IN}$ and $\widetilde{\mathcal{T}}_\text{IN}$.

We now make the following mild restriction on protocols allowed: 
\begin{equation}
\label{eq:restriction}
\widetilde{w}^0_\text{IN}=\widetilde{w}^0.
\end{equation}
We say this is mild, because the trajectories
$\widetilde{\mathcal{T}}_\text{OUT}$ have an extra work {\em gain} relative to
their sister trajectories in $\mathcal{T}_\text{OUT}$ following from their
initial lowering. This gain tends to infinity as $p(OUT)\rightarrow
0$.  Thus the $\sim$-protocol will not have worse possible work values than the actual protocol. The sort of protocols that are ruled out by the mild assumption are those where there is initially a dramatic quench without thermal hopping such that one of the OUT levels is raised very high. As long as it is not raised above the initial value of the level in the $\sim$-protocol the mild assumption is not violated however---the shifting down and up of the level would have a net work gain. This way the mild assumption does not rule out e.g.\ 2-level Szilard engine protocols where the second, likely empty, level is lifted very high initially and then lowered quasistatically. This example is studied in a physical system in Appendix~\ref{appendix:B} below.  Moreover,  protocols involving thermalisation at the start, including quasistatic protocols, or appropriately bounded initial quenches, are not ruled out.


Combining Eqs.\ref{eq:worstcaseequiv} and \ref{eq:restriction} gives the desired expression used in the main body:
$$w^0_\text{IN}=\widetilde{w}^0.$$

\begin{figure}
\centering
\includegraphics[width=0.6\linewidth]{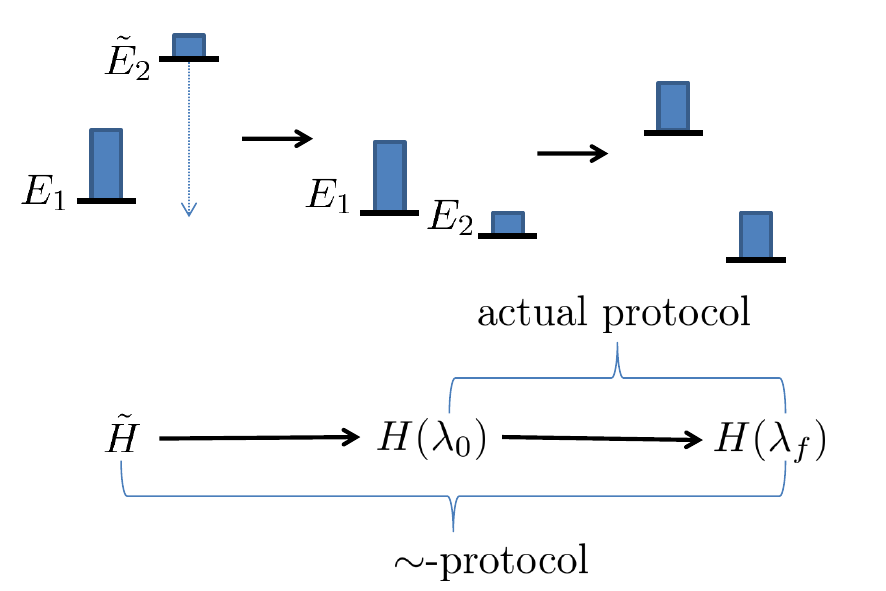}
\caption{A very simple example of how we prepend an extra step to the actual protocol as part of the theoretical analysis. Here level 2 is designated as OUT. The energy of that level is initially higher than in the actual initial Hamiltonian $H(\lambda_0)$ so that its occupation probability is thermal. Then it is lowered down without interacting with a thermal bath. Now the actual protocol begins. 
}
\label{fig:simpleexample}
\end{figure}

To help illustrate the notation a simple example of the $\sim$-protocol and how it relates to the actual protocol is in Figure~\ref{fig:simpleexample}.

\section{Smooth relative entropy}\label{app.smooth}
As noted in the main body, the optimum term reduces to a relative entropy in a special case.
If $H(\lambda_0)=H(\lambda_f)$, 
$\log  \left(Z/\widetilde{Z}\right)=-\log \sum_{i\in supp(\rho_0)}
e^{-\beta E_i}/Z_f=D_0(\rho_0||\gamma)$. Moreover if 
$H(\lambda_0)=H(\lambda_f)=0$,  $D_0(\rho_0||\gamma)=\log
d-S_{\max}(\rho_0)$ (noting $\gamma=\id/d$). This recovers the known
results from~\cite{DahlstenRRV11,HorodeckiO13,Aberg13} that these are optimal in the respective cases. If
$p(OUT)$ defined above is not necessarily zero, this optimal term depends on
which levels are chosen to be in $\mathcal{E}_{OUT}$. If one chooses the {\em best} cut
between IN and OUT, in the sense of minimizing $\widetilde{Z}$ and thus the
worst-case work, the optimum one can hope for becomes in those cases
$-D_0^{\epsilon}(\rho_0||\gamma):=\min -D_0(\rho'||\gamma)$
such that $d(\rho_0,\rho')\leq \epsilon$ where
$d$ is the 
trace 
distance (this is 
called the {\em smooth} relative entropy). The interpretation is that
the optimal worst-case work one can hope for allowing for an error tolerance of
$\epsilon=p(OUT)$ is $kT D_0^{\epsilon}(\rho_0||\gamma)$, consistent
with~\cite{DahlstenRRV11,HorodeckiO13,Aberg13}.

\section{Relation between worst-case and deterministic work}
Certain protocols studied in single-shot statistical mechanics give a pseudo-deterministic work output, i.e. the work distribution is highly concentrated around some value. (It is standard to say that a certain amount of work $A$ is ($\delta$,$\epsilon$)-deterministic if one will have $W\in \left[ A\pm\delta \right] $ except with probability $\leq \epsilon$. ) For example in~\cite{DahlstenRRV11} one may compress all the randomness onto some bits and extract work from the others with essentially deterministic work output. In ~\cite{HorodeckiO13,Aberg13} and and~\cite{SalekW16} the expressions given concern the optimal pseudo-deterministic work, optimised over protocols for some given boundary conditions. This is in contrast to this paper and e.g.~\cite{EgloffDRV12} which make bounds on the optimal worst-case work. We note that from the definitions one sees that bounds on worst-case work are also bounds on deterministic work but not vice versa. This is because demanding that the work cost is bounded from above is a weaker restriction than demanding that it is bounded from both above and below.

\section{Cutting the work-tail, as well as the state-tail}
There can actually be (sets of) trajectories which are unlikely even if the initial state of the trajectory is likely, as the hopping probability may be low. For example, if one lifts one level towards a very high value while thermalizing, there is one trajectory corresponding to staying in the rising level throughout, which would be the trajectory that gives the worst-case work. However, if such a trajectory is very unlikely, one would wish to ignore the trajectory when stating the worst-case work. In this section, we show a way to ignore such unlikely trajectories, by not only cutting off part of the initial state as previously, but also cutting off a part of the work distribution. This strategy gives a different penalty term---lower, in general---in the equality for the worst-case work. 

\noindent {\bf {\em Proof overview}---}We shall again take the initial density matrix to have the form
$\rho_0= \sum_{i=1}^{d}p_i\ket{i_0,\lambda_0}\bra{i_0,\lambda_0}$, not
necessarily a thermal state. Then a sequence of Hamiltonian changes
and thermalizations as described above is applied. This induces some
work probability distribution and some worst-case work for the trajectories of interest.

The argument is split in two. First, we define a set of trajectories
of interest: Some trajectories are unlikely enough to be ignorable. We
derive the worst-case work for the set of trajectories of interest. Next, we consider the probability that some trajectory is in the set of interest. Combining these two parts gives our new equality for worst-case work.

\noindent {\bf{\em The set of trajectories of interest}}---We wish to ignore unlikely trajectories. We identify a set of trajectories of interest, defined as excluding trajectories of two types:
 
{\bf 1.  $\rho_0$-tail trajectories}: These are those which are called $\mathcal{T}_{OUT}$ above, i.e., trajectories which start in $\mathcal{E}_{OUT}$. We now call them $\rho_0$-tail trajectories as using $OUT$ risks generating confusion because of the second type of cut we shall make on the set of trajectories. 

 {\bf 2. Work-tail trajectories}: We also ignore trajectories associated with the worst work values, if those values are sufficiently improbable. 
This ignoring amounts to cutting off the worst-case tail of the work probability distribution.
To simplify the proof, we define this tail in terms of the work probability distribution of the fictional thermal state $\widetilde{\gamma}$. By ``the work-tail,'' we mean the set of trajectories associated with the following work values $w$: If the initial state is $\widetilde{\gamma}$, there is an associated work probability distribution $\widetilde{p}_{\fwd}(w)$ for the given protocol, 
 and an associated worst-case work $\widetilde{w}^\epsilon$. The work tail trajectories are by definition those with work cost $w>\widetilde{w}^\epsilon$.
 Since the actual initial state $\rho_0$ may differ from $\widetilde{\gamma}$, the probability that some trajectory begins in the work tail does not necessarily equal $\epsilon$.

\begin{figure}
\centering
\includegraphics[width=0.4\linewidth]{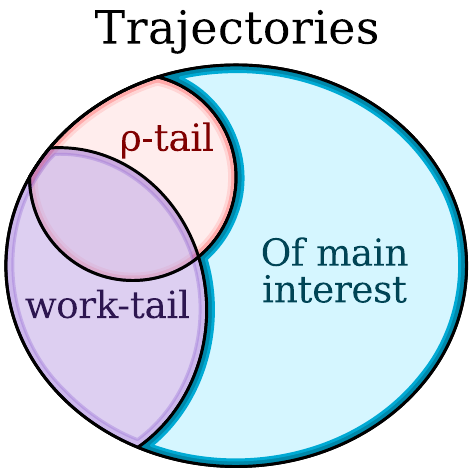}
\caption{Depiction of the trajectories of interest. We shall ignore trajectories that have undesirable, very unlikely work values (that are in the work-tail) and trajectories that start in very unlikely energy eigenstates (that start in the $\rho$-tail).}
\label{fig:trajectories}
\end{figure}

 These sets are depicted in Fig. \ref{fig:trajectories}. We shall call the worst-case work in the set of interest $w^0_{IN,IN}$.

\noindent {\bf {\em The worst-case work in that set}---}We now derive the worst-case work in the set of trajectories of interest: We maximize the work cost $w$ over that set of interest.
We shall, for the first part, draw inspiration from an argument, in~\cite{YungerHalpernGDV14}, concerning scenarios governed by Crooks' Theorem. Take the initial state of the forwards process to be $\rho_0=\widetilde{\gamma}$; and the initial state of the reverse process as $\gamma=e^{-\beta H(\lambda_f)}/Z_f$. 

Maximize Crooks' Theorem over the support of $\tilde{p}_{\fwd}(w)$~\cite{YungerHalpernGDV14}:
$$\max \frac{\widetilde{p}_{\fwd}(w)}{p_{\rev}(-w)}=\max \frac{Z}{\widetilde{Z}} e^{ \beta w}.$$
The RHS is monotonic in $w$, so maximizing the RHS over the support of 
$\tilde{p}_{\fwd}(w)$ leads to the maximum $w$-value $w^0$. Taking the logarithm and recalling the $D_\infty$ definition yields~\cite{YungerHalpernGDV14}, 
$$\beta w^0 = D_{\infty} (\widetilde{p}_{\rm fwd}(w) || p_{\rm rev}(-w))-\log\left(\frac{Z}{\widetilde{Z}}\right).  $$

Now, we cut off the work tail by defining a cut-off probability distribution
$   \widetilde{p}_{\fwd}^{\epsilon}(w)
   :=   0$,  if  $w\geq \widetilde{w}^{\epsilon}$ and
        $\frac{  \widetilde{p}_{\fwd}(w)  }{  1-\epsilon  }$, otherwise, wherein $\widetilde{w}^{\epsilon}$ denotes the work guaranteed up to probability $\epsilon$ if $\widetilde{\gamma}$ is the initial state.
[Dividing by ($1-\epsilon$) normalizes the distribution.]
For work values outside the work tail, Crooks' Theorem can be reformulated as
$$\frac{\widetilde{p}_{\fwd}^{\epsilon}(w)}{p_{\rev}(-w)}(1-\epsilon)=\frac{Z}{\widetilde{Z}} e^{ \beta w}.$$

Since the RHS is monotonic,
$$\max \frac{\widetilde{p}_{\fwd}^{\epsilon}(w)}{\widetilde{p}_{\rev}(-w)}(1-\epsilon)=\left(\frac{Z}{\widetilde{Z}}\right) e^{\beta \widetilde{w}^{\epsilon}},$$
wherein the maximization is over the support of $\widetilde{p}_{\fwd}^{\epsilon}$. Taking the logarithm and rearranging yields
\[
 \beta \widetilde{w}^{\epsilon} 
\!=\! D_{\infty} (\widetilde{p}_{\fwd}^{\epsilon}(w) \!|| \widetilde{p}_{\rev}(-w)) 
+\log(\!1\!-\!\epsilon)
-\log  \left( \frac{Z}{\widetilde{Z}} \right).
\] 
The LHS is the worst-case work in the set of trajectories of interest.

\noindent {\bf {\em Probability that a trajectory is in the set of interest}---}The trajectories of interest are effectively the possible trajectories. To make precise what is meant by ``effective,'' we bound the probability that any particular trajectory lies outside that set.

Consider a trajectory followed by a system initialized to $\rho_0$. The probability that the trajectory lies outside the set of interest is bounded by $p(\rho_0{\rm -tail})+p({\rm work\!-\!tail})$, as shown in Fig.~\ref{fig:trajectories}. $p(\rho_0 {\rm-tail})$, defined via $\rho_0$ and the choice of effective support, is specified by input parameters. $p({\rm work\!-\!tail})$ denotes the probability that the trajectory is in the set associated with a worse work cost than $\widetilde{w}^{\epsilon}$ (the work guaranteed up to probability $\epsilon$ not to be exceeded, 
if the initial state is $\widetilde{\gamma}$). $p({\rm work\!-\!tail})$ does not necessarily equal $\epsilon$ for an arbitrary $\rho_0$. As $p({\rm work\!-\!tail})$ is not an input parameter, we wish to bound it with input parameters. 
 
Let us drop the subscript ``fwd'' and refer simply to $p(w)$. 
The weight $p(w>x)$ in the actual work tail associated with $\rho_0$ cannot differ arbitrarily from the weight  
$\widetilde{p}(w>x)$ in the work tail associated with $\widetilde{\gamma}$: 
$$|p(w>x)-\widetilde{p}(w>x)|\leq   d \Large( p(w),\widetilde{p}(w) \Large).$$
This bound follows from the definition of the variational distance $d$, which equals the trace distance between diagonal states.\footnote{See, e.g., Sec. 2 in \url{http://people.csail.mit.edu/costis/6896sp11/lec3s.pdf.}}

The variational distance $d$ is contractive under stochastic matrices,
because the trace distance is contractive under completely positive
trace-preserving (CPTP) maps. The work distribution is
the result of a stochastic matrix's acting on the probability
distribution over initial energy eigenstates.  Let us now in this
paragraph for convenience use Dirac notation for classical probability
vectors, representing a probability distribution $p(w)$ as $\langle
w\ket{p}$.
The 
work distribution comes from the stochastic matrix
$\sum_j\ket{p_j}\bra{j}$ 
mapping a state $\ket{\rho_0}$ to a
work distribution, wherein $j$ labels
projectors onto $H(\lambda_0)$ eigenstates, $\ket{p_j}$ labels the work
distribution when starting with an initial state $\ket{j}$ 
(i.e.,~$p_j(w)=\langle w|p_j\rangle$), and $\ket{\rho_0}=\sum_j
q_j\ket{j}$. For example, if there are two possible eigenstates, we
can write $\ket{\rho_0}=q_1\ket{1}+q_2\ket{2}=(q_1\,\,q_2)^T$, and the
resulting work distribution
$p(w)=(\langle w\ket{p_1}\bra{1}+\langle
w\ket{p_2}\bra{2})\ket{\rho_0}=q_1 p_1(w)+q_2 p_2(w)$. 

Thus,
$$|p(w>x)-\widetilde{p}(w>x)|\leq   d \Large( p(w),\widetilde{p}(w) \Large) \leq d (\rho_0,  \widetilde{\gamma}) \, \forall x. $$ 
For some $x=x'$,  by definition, $\widetilde{p}(w>x')=\widetilde{p}({\rm work\!-\!tail})=\epsilon$, and $p({\rm work\!-\!tail}):=p(w>x')$. Thus 
$$p({\rm work\!-\!tail})\leq   d (\rho_0,\widetilde{\gamma})   +\epsilon.$$

\noindent {\bf {\em Main result, also cutting work tail}---}We conclude that the worst-case work from the trajectories of interest, $w^0_{IN,IN}$ respects
\begin{align}  \label{eq:MainResult}
 \beta w^0_{IN,IN} 
\!=\! D_{\infty} (\widetilde{p}_{\fwd}^{\epsilon}(w) \!|| p_{\rev}(-w)) 
+\log(\!1\!-\!\epsilon)
-\log Z/\widetilde{Z} .
\end{align}

 The probability that the trajectory is not in the set of interest is upper-bounded by  $ p(\rho_0\text{-}tail)+p(work\text{-}tail)   
\leq   p( \rho_0  \text{-}tail)   +   d ( \rho_0,  \:  \widetilde{\gamma})  +\epsilon$.

\section{Continuous time versus discrete time} 
\label{section:Continuous}

We have mainly focused on the discrete-time protocol. Experimental
realizations of thermodynamic protocols are often described by a
continuous master equation. Here, we show that the discrete protocol
leads to a master equation in the continuum model and vice versa. In
this section we restrict ourselves to scenarios without energy
coherences, i.e., the discrete-classical case.  

\subsection{From discrete to continuous}
We consider a discrete sequence of times, $t_m = t_0 + m \: dt$ ($m=0,1,2  \ldots$), and the sequence $\lambda_m\equiv\lambda(t_m)$ of values of the external parameter. As the waiting time decreases ($dt\to0$), the transition probability
$p(\ket{i,\lambda(t),t}\to\ket{j,\lambda(t+dt),t+dt})$ due to thermalization should vanish. To  first order, it behaves as
\begin{equation}
p(\ket{i,\lambda(t),t}\to\ket{j,\lambda(t+dt),t+dt})
\approx \delta_{ij} + \Gamma_{i\to j}(t)dt
+ \mathcal{O}(dt^2).
\end{equation}
The transition rate $\Gamma_{i\to j}(t)$ is a possibly complicated function of instantaneous energy levels $E(\ket{i,\lambda(t),t})$. However, the transition rates inherit the condition
\begin{equation}
\frac{\Gamma_{i\to j}(t)}{\Gamma_{j\to i}(t)}
= e^{-\beta\left[
E(\ket{j,\lambda(t),t})-E(\ket{i,\lambda(t),t})
\right]}
\end{equation}
from detailed balance and the condition
\begin{equation}
\sum_{j}\Gamma_{i\to j}(t) = 0
\end{equation}
from probability conservation.
The occupation probability is
\begin{align*}
p(\ket{j,\lambda(t+dt),t+dt})
&= \sum_ip(i,\ket{\lambda(t),t})
p(\ket{i,\lambda(t),t}\to\ket{j,\lambda(t+dt),t+dt}) \\
&\approx p(\ket{j,\lambda(t),t})
  + \sum_ip(i,\ket{\lambda(t),t})\Gamma_{i\to j}(t)\,dt
  - \sum_ip(j,\ket{\lambda(t),t})\Gamma_{j\to i}(t)\,dt.
\end{align*}
If the occupation probability is a smooth function of time, the master equation
\begin{equation}
\frac{d}{dt}p(\ket{j,\lambda(t+dt),t+dt})
= \sum_ip(i,\ket{\lambda(t),t})\Gamma_{i\to j}(t)
- \sum_ip(j,\ket{\lambda(t),t})\Gamma_{j\to i}(t)
\end{equation}
follows.
The equivalence is further illustrated in Appendix~\ref{appendix:B} in the example of an electron box.

\subsection{From continuous to discrete}
Going in the other direction, we now show explicitly how the discrete-time
model can be derived from a physical master equation.  Consider a
two-level system that has a state $|0\rangle$, kept at zero 
energy, and a state $|1\rangle$ whose energy $\hbar\omega(t)$
changes. The Hamiltonian is $H(t)=\hbar\omega(t)|1\rangle\langle 1|$,
and the system interacts with a temperature-$T$ heat bath.
In~\cite{albash_quantum_2012}, a master equation for the 
density matrix $\rho(t)$ was derived for a such system. In
the present case, the master equation is
\begin{eqnarray}
  \label{eq:master}
  \dot{\rho}(t)&=&-i[H(t),\rho(t)]+{\cal L}(t)\rho(t)\\
{\cal L}(t)\rho&=&\Gamma d(\omega(t))
\left(
[n_{\rm th}(\omega(t))+1]   \left\{ [\sigma_-,\rho(t)\sigma_+]+h.c.\right\}
+n_{\rm th}(\omega(t))\left\{ [\sigma_+,\rho(t)\sigma_-]+h.c.\right\}
\right).
\end{eqnarray}
The heat bath, modeled as as set of harmonic oscillators, 
has a thermal occupation number 
$n_{\rm th}(\omega)=(e^{\beta  \hbar\omega}-1)^{-1}$
that depends on time
because the upper level shifts. $d(\omega)$ is the dimensionless
heat-bath density of states; $\Gamma$ denotes a rate assumed to be
constant; 
$\sigma_-=|0\rangle\langle 1|$ denotes the usual lowering operator; and
$\sigma_+=\sigma_-^\dagger$.  Equation~(\ref{eq:master}) has the form of
the usual Lindblad master equation, but the Lindblad operator depends on time. The dependence arises only from the level spacing's time dependence. 
The Hamiltonian part contains the Lamb shift.

In the derivation of Eq.~(\ref{eq:master}) one assumes, as usual, weak coupling to the heat bath, the Markovian approximation, and the rotating-wave approximation.  One also assumes that the adiabatic approximation holds, i.e., the system always remains in its time-local energy eigenstates when the interaction with the heat bath is ignored. This condition is always fullfilled under the assumption of vanishing energy coherences at all times that we made in this
section. Indeed, the part of (\ref{eq:master})  pertaining to the
diagonal elements of $\rho(t)$ can be derived without the
adiabatic assumption~\cite{Ingold92a}. 

We now consider discrete times $t_n:= n \Delta t $, $n=0,\ldots,
N$, with $\omega(t)$ constant during the time intervals $\Delta t$,
$\omega_n  :=  \omega(t_n)$. Restricting ourselves to changes of
the Hamiltonian that only involve its spectrum, $H(t)$ and ${\cal
  L}(t)$ are constant during a given time interval. 

Consider first the Hamiltonian changes. 
Heisenberg's equation of motion for the system-and-bath composite
implies that $\dot{\rho}(t)$ has a finite jump when the Hamiltonian 
has a finite jump.  Therefore, $\rho(t)$ is continuous 
when the Hamiltonian has a finite jump. Hence for finite
Hamiltonian changes during a time $\delta t$, the system-and-bath
composite's density matrix is unchanged in the limit as $\delta t\to
0$. Hence the system's reduced density matrix is unchanged during the
instantaneous shift of energy levels.  
As for the relaxation process, the initial thermal state is described
in terms of occupation 
probabilities $p_n$ for the $n^{\rm th}$ level. The evolution during the
relaxation process is given by $p(t)=e^{Tt}p(0)$, where $T$ is a
matrix that connects the diagonal matrix elements of $\rho$ in the
master equation (\ref{eq:master}):
$\dot{\rho}_{nn}=\sum_mT_{nm}\rho_{mm}$.  The transition rates
$T_{nm}$ inherit detailed balance from
the rates 
appearing in the master equation,
i.e.,~$T_{ij}=e^{-\beta(\epsilon_i-\epsilon_j)}T_{ji}$.  
Detailed balance holds for each power $T^k$ of $T$,
$(T^k)_{ij}=e^{-\beta(\epsilon_i-\epsilon_j)}(T^k)_{ji}$, for all
$k\in\mathbb{N}$.
Therefore, by the power-series expansion of $e^{Tt}$,
detailed balance holds also for $e^{Tt}$.
We thus have derived, from a
physical model of a system that is coupled to a heat bath and whose
energy levels 
are piecewise-constant, the discrete-time model considered in the paper.\\

To illustrate this let us consider a two level system:
Expressing $\rho(t)   =   p_0(t)|0\rangle\langle 0|   +   [1-p_0(t)]   |1\rangle
\langle 1|$, we obtain a differential equation for $p_0(t)$,
\begin{equation}
  \label{eq:pdot2}
  \dot{p}_0(t)+   g(\omega(t))   p_0(t)
  =   \frac{g(\omega(t))}{2}+\Gamma d(\omega(t)),
\end{equation}
wherein $g(\omega(t))  :=  
2\Gamma   d ( \omega(t) )   [2 n_{\rm th}(\omega(t))+1]$. 
This equation has the general solution 
\begin{eqnarray}
  \label{eq:pt}
  p_0(t)&=&\left(p_0(0)-\frac{1}{2} +\Gamma\int_0^tdt_1 d(\omega(t_1))G(t_1)
  \right)/G(t)+\frac{1}{2},
\end{eqnarray}
wherein $G(t)  :=  e^{\int_0^{t}g(\omega(t_1))dt_1}$. 
The integrals in 
Eq.~(\ref{eq:pt}) can be calculated analytically:
\begin{equation}
  \label{eq:p0const}
  p_0(t)=p_0(0){\rm e}^{-2
    d(\omega)\coth(\frac{\beta\hbar\omega}{2})\Gamma t}+ p_{0,{\rm
      th}}   \left[   1-{\rm e}^{-2
    d(\omega)\coth(\frac{\beta\hbar\omega}{2})\Gamma t}   \right],
\end{equation}
wherein $p_{0,{\rm th}}   :=   1/(e^{-\beta\hbar\omega}+1)$ denotes
the ground state's thermal 
occupation. For large times, the memory
of the initial state is lost, and the system relaxes towards thermal
equilibrium.  From Eq.~(\ref{eq:p0const}), we obtain
the transition probabilities during relaxation over the time
interval between $ n\Delta t$ and $(n+1)\Delta t$:
$p(|0_n,\omega_n\rangle\rightarrow |0_{n+1},\omega_n\rangle)
=p_0(\Delta t)|_{p_0(0)=1}$, $p(|0_n,\omega_n\rangle\rightarrow
|1_{n+1},\omega_n\rangle)= 1- p(|0_n,\omega_n\rangle\rightarrow
|0_{n+1},\omega_n\rangle)$, $p(|1_n,\omega_n\rangle\rightarrow
|0_{n+1},\omega_n\rangle) =p_0(\Delta t)|_{p_0(0)=0}$, and
$p(|1_n,\omega_n\rangle\rightarrow |1_{n+1},\omega_n\rangle)= 1-
p(|1_n,\omega_n\rangle\rightarrow |0_{n+1},\omega_n\rangle)$.  
These transition 
probabilities obey detailed balance.  As they remain unchanged by the
inclusion of an instantaneous Hamiltonian change at the end of each
time interval, 
$p(|i_n,\omega_n\rangle\rightarrow
|j_{n+1},\omega_n\rangle)=p(|i_n,\omega_n\rangle\rightarrow
|j_{n+1},\omega_{n+1}\rangle)$ for $i,j\in\{0,1\}$. \\

\section{Application to solid-state system: Electron box}
\label{appendix:B}

To demonstrate the physical relevance of our results, we take a realistic example, the  electron box~\cite{Koski14a,Koski13a,Saira12a,Altshuler91a,Grabert92a}, and apply our results to it.
We first derive a time-local master equation for the level-occupation probabilities in Appendix~\ref{appendix:B.1}. As shown in Appendix~\ref{section:Continuous}, 
the master equation is equivalent to the discrete-time trajectory model discussed in the main text. The work distribution functions are analyzed numerically in Appendix~\ref{appendix:B.2} and analytically in Appendix~\ref{GuaranteedW::sec:1}. 
Finally, we upper-bound the penalty term $D_\infty$, which reveals the direct physical relevance of our results.

\subsection{Theoretical model and its justification}
\label{appendix:B.1}

\begin{figure}
\centering
\includegraphics[width=5cm]{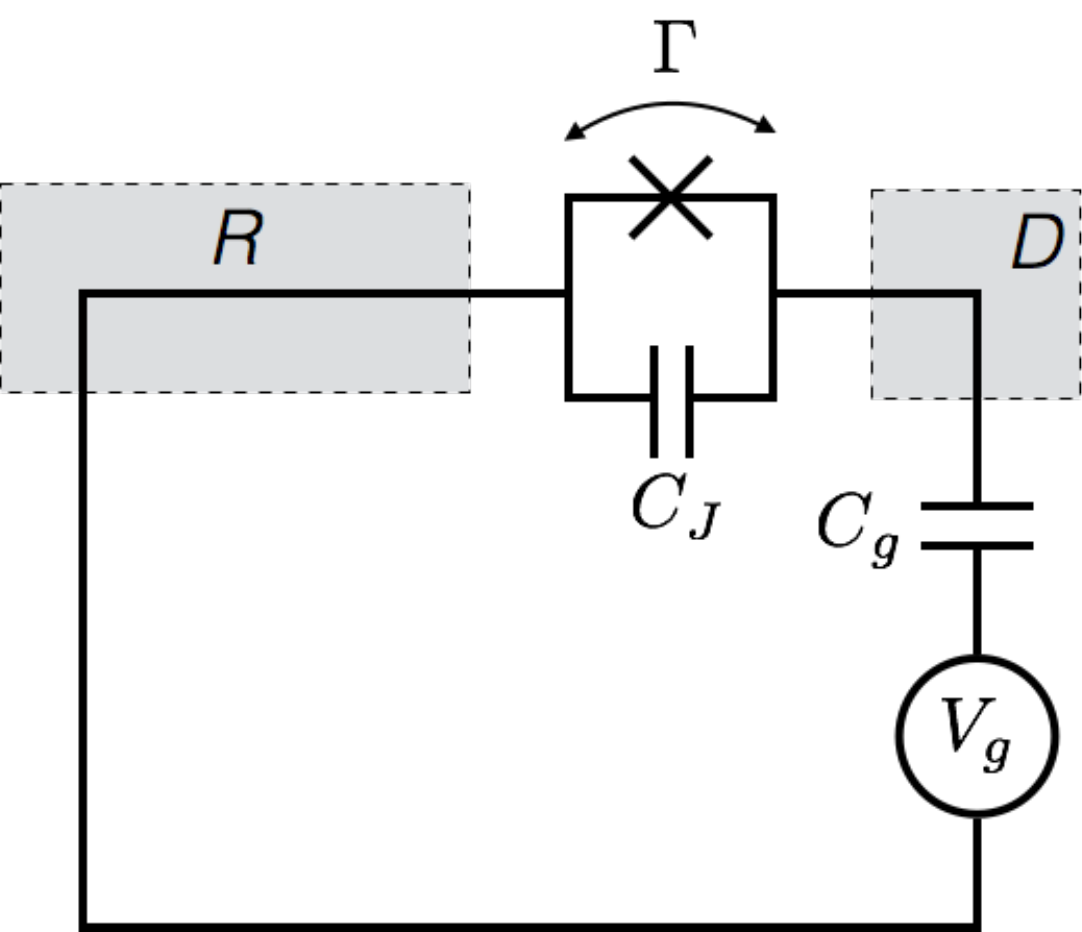}
\caption{A schematic of an electron box.}
\label{WorkDist::fig:6}
\end{figure}

We consider the type of system in~\cite{Koski14a, Koski13a, Saira12a}.  Following a semiclassical theory (known as ``the orthodox theory'') such as in
\cite{Amman91a}, we derive a master equation and illustrate the work fluctuations.
A more complete quantum description is possible~\cite{Altshuler91a,Grabert92a}. Yet the semiclassical approach is useful for interpreting and identifying work and heat, which are often ambiguous.

The system (Fig.~\ref{WorkDist::fig:6}) consists of a large metallic electrode $R$ that serves as a charge reservoir, a small metallic island (or quantum dot) $D$, and a gate electrode. The island $D$ is coupled only capacitively to the gate electrode but couples to the reservoir $R$ capacitively and via tunnelling. The Hamiltonian has four parts:
\begin{math}
H = H_R + H_{D} + H_C + H_T.
\end{math}
The first two terms,
\begin{equation}
H_R = \sum_k \varepsilon_k c_k^\dag c_k 
\quad {\rm and} \quad
H_D = \sum_q\varepsilon_q d_q^\dag d_q,
\end{equation}
describe the non-interacting parts of the electrode $R$ and the island $D$. Here, $c_k^\dag$ ($d_q^\dag$) creates an electron with momentum $\hbar k$ ($\hbar q$) and energy $\varepsilon_k$ ($\varepsilon_q$). The single-particle dispersions $\varepsilon_k$ and $\varepsilon_q$ form continua of energy levels.
$H_C$ is responsible for the electron-electron interaction on the island. We adopt a capacitive model as recounted below.
Finally, the tunnelling of electrons between $R$ and $D$ is described by
\begin{equation}
H_T = \eta\sum_{kq} c_k^\dag d_q + h.c.,
\end{equation}
wherein $\eta$ is the tunnelling amplitude.
$\eta$ is assumed not to depend on momenta (or on energy), as in common metals that have wide conduction bands.

The effective semiclassical model describes equilibrium:
Suppose that an electron tunnels between the island $D$ and the reservoir $R$.
The tunneling jolts the system out of equilibrium, but the system equilibrates quickly:
Coulomb repulsions redistribute the electrons throughout the circuit.
After the redistribution,
the junction carries the equilibrium charge $Q_J$, 
and the gate carries the equilibrium charge $Q_g$.
These charges are regarded as being ``on'' the island $D$, 
due to the island's capacitive couplings to the reservoir $R$ and to the gate.
(The island carries also \emph{excess} electrons, discussed below.)
The electrons continue to repel each other,
imbuing the system with the equilibrium Coulomb energy
\begin{equation}
\label{WorkDist::eq:6}
H_C = \frac{Q_J^2}{2C_J}
+ \frac{Q_g^2}{2C_g},
\end{equation}
wherein $C_J$ and $C_g$ denote the junction and gate capacitances.
One can find that
\begin{subequations}
\label{WorkDist::eq:17}
\begin{align}
  Q_J & = C(V_g - Ne/C_g) \,,\\
  Q_g & = C(V_g + Ne/C_J) \,,
\end{align}
\end{subequations}
wherein $C := C_gC_J/(C_g+C_J)$ is the system's effective capacitance. 
$N=\sum_k d_k^\dag d_k$ denotes the number of \emph{excess} electrons,
relative to the charge-neutral state, on the island $D$.
When $N = 0$, the island has zero net charge.
$H_C$ can thus be rewritten as
\begin{equation}
H_C = E_C N^2 + \frac{1}{2} CV_g^2,
\end{equation}
wherein $E_C:= e^2/2(C_g+C_J)$ is the single-electron \emph{charging energy}, one of the largest energy scales of the system.

We are primarily interested in the macroscopic variable $N$ but not in the microscopic degrees of freedom $c_k$ and $d_q$, whose dynamics is typically much faster. One can thus integrate out $c_k$ and $d_q$ to get the effective Hamiltonian expressed only in terms of $N$. In the semiclassical approach, this can be achieved by considering the energy that an electron gains by tunnelling.

Suppose that an electron tunnels into the island $D$ from the reservoir $R$.
This will change the charge $Q_J\to Q_J-e$ and the excess number of electrons $N\to N+1$. This new charge configuration, right after the tunnelling, is redistributed quickly to a new equilibrium configuration,
\begin{subequations}
\begin{align}
  Q_J' & = C[V_g - (N+1)e/C_g] \\
  Q_g' & = C[V_g + (N+1)e/C_J] ,
\end{align}
\end{subequations}
by the gate voltage source. The voltage source has moved the amount 
\begin{equation}
\Delta{Q} := Q_J'-(Q_{ J }  -  e) = eC_g/(C_g+C_J)
\end{equation}
of charge through the transmission line from the junction interface to the gate capacitor by doing the amount
\begin{equation}
W = V_g\Delta{Q}
= eV_gC_g/(C_g+C_J)
\end{equation}
of work on the system.
Therefore, the electron's overall energy gain $\Delta{E}$ equals the work $W$ minus the change in the electrostatic energy:
\begin{equation}
\Delta{E} = E_C\left[2C_gV_g/e - (2N+1)\right] \,.
\end{equation}
As this energy gain comes from the transition $N\to N+1$, the effective Hamiltonian for the macroscopic variable $N$ can be regarded as
\begin{equation}
\label{WorkDist::eq:19}
H_\text{eff} = E_C(N^2 - 2NN_g),
\end{equation}
wherein $N_g:= C_gV_g/e$. Recall that the second term comes from the work done on the system by the voltage source.

The microscopic degrees of freedom 
removed from the effective macroscopic model
cause $N$ to fluctuate randomly.
The transition $N\to N\pm1$ is associated with tunneling of an electron 
into or from the island.
Hence the transition rate follows from Fermi's Golden Rule:
\begin{equation}
\label{WorkDist::eq:20}
\Gamma(\Delta{E}) \approx
\frac{2\pi|\eta|^2\rho_R\rho_D}{\hbar}
\frac{\Delta{E}}{e^{\beta\Delta{E}}+1} \,,
\end{equation}
wherein $\rho_R$ and $\rho_D$ are the densities of states of $R$ and $D$, respectively, and
\begin{equation}
\Delta{E} = H_\text{eff}(N\pm 1)-H_\text{eff}(N) \,.
\end{equation}

Finally, at sufficiently low temperatures ($\beta E_C\gg 1$), higher energy levels play no role. Considering the two lowest levels $N=0$ and $N=1$ suffices for $N_g\in[0,1]$.\footnote{The model is invariant under $N_g\to N_g+1$, and studying $N_g\in[0,1]$ suffices.} 
Let $p_0$ denote the probability that $N = 0$, 
and let $p_1$ denote the probability that $N = 1$.
With Eqs.~(\ref{WorkDist::eq:19}) and (\ref{WorkDist::eq:20}), 
this two-level approximation leads to the master equation
\begin{subequations}
\label{WorkDist::eq:4}
\begin{align}
\dot{p}_0 & = -\Gamma_+p_0 + \Gamma_-p_1 \\
\dot{p}_1 & = -\Gamma_-p_1 + \Gamma_+p_0.
\end{align}
\end{subequations}
The transition rates are ~\cite{Altshuler91a, Grabert92a}
\begin{equation}
\Gamma_\pm(t) :=
\Gamma(\pm\epsilon(t)) 
\quad  {\rm and} \quad
\Gamma(\epsilon) :=
\frac{\Gamma_0\epsilon(t)/\varepsilon_c}{e^{\beta\epsilon(t)}-1}.
\end{equation}
Here, $\varepsilon_c$ is the bath's high-frequency cutoff (i.e., $\hbar/\varepsilon_c$ is the correlation time), and $\Gamma_0$ is a constant that characterizes the strength of the coupling to the bath.
$\Gamma_0/\varepsilon_c$ is related to the material properties by $\Gamma_0/\varepsilon_c=2\pi|\eta|^2\rho_R\rho_D/\hbar$. Note that the transition rates satisfy the detailed-valance relation
\begin{equation}
\label{WorkDist::eq:15}
\frac{\Gamma(+\epsilon)}{\Gamma(-\epsilon)}
= e^{-\beta\epsilon}.
\end{equation}
The time-local master equation~(\ref{WorkDist::eq:4}) is equivalent to the discrete-time trajectory model (see Appendix~\ref{section:Continuous}). Therefore, the electron box is a realistic prototype system to which our results can apply.

\subsection{Monte Carlo simulation of the electron box}
\label{appendix:B.2}

We performed a Monte Carlo simulation of an erasure protocol in the electron box set-up.
Our simulation discretizes the protocol into time steps $\delta t$ small enough to justify the linear approximation that the population of level $i$ evolves from time step $t$ to $t+\delta t$ according to $p_i\left(t+\delta t\right)=p_i\left(t\right)+\delta t \dot{p}_i\left(t\right)$.
Using Eqs.~(\ref{WorkDist::eq:4}), we can write a stochastic matrix acting on the probabilities:
\begin{equation}
\label{eq:StochasticEB}
\left[\begin{array}{c}
p_0   \left(t + \delta t\right) \\
p_1    \left(t + \delta t\right) 
\end{array}\right]
=
\left[\begin{array}{cc}
1 - \Gamma_+\delta t & \Gamma_-\delta t \\
\delta t \Gamma_+ & 1 - \Gamma_-\delta t \\
\end{array}\right]
\left[\begin{array}{c}
p_0    \left(t\right) \\
p_1    \left(t\right) 
\end{array}\right].
\end{equation}

For a two-level system which does not build up quantum coherences, a stochastic thermalizing matrix (which by its definition evolves all states towards the Gibbs state) has only one degree of freedom remaining once the Gibbs state has been chosen: 
the thermalization speed. All models of two-level thermalizations for a given Gibbs state are equivalent.
We pick the conceptually straightforward {\em partial swap}:
With some probability $p_\mathrm{sw}$, the system's current state is exchanged with the Gibbs state. With probability $1 - p_\mathrm{sw}$, 
the state remains unchanged:
$M_\mathrm{swap} = (1-p_\mathrm{sw})\id + p_\mathrm{sw} \ketbra{\mathrm{Gibbs}}{\mathrm{ones}}$, where $\ket{ {\rm ones} }$ means the vector of 1's. 
For a Gibbs state associated with an energy-level splitting $\epsilon$, 
\begin{equation}
\label{eq:StochasticPS}
M_\mathrm{swap}  =  \left[\begin{array}{cc}
1 - \dfrac{p_\mathrm{sw} \exp\left(-\beta\epsilon\right)}{1+ \exp\left(-\beta\epsilon\right)} & \dfrac{p_\mathrm{sw}}{1+\exp(-\beta\epsilon)} \\
\dfrac{p_\mathrm{sw} \exp\left(-\beta\epsilon\right)}{1+\exp(-\beta\epsilon)} & 1 - p_\mathrm{sw}\dfrac{1 - \exp\left(-\beta\epsilon\right)}{1 + \exp\left(-\beta\epsilon\right)}
\end{array}\right].
\end{equation}

Equating Eq.~(\ref{eq:StochasticPS}) with the matrix in Eq.~(\ref{eq:StochasticEB}),
we can find the partial-swap probability in terms of 
the electron box's physical parameters: 
\begin{equation}
\label{eq:SwapProbEB}
p_\mathrm{sw}\left(t\right) = \dfrac{\Gamma_0 \delta t }{\varepsilon_c} \epsilon\!\left(t\right) \coth\left[\beta\epsilon\!\left(t\right)/2\right].
\end{equation}
The swap probability $p_\mathrm{sw}\left(t\right)$ and the energy level splitting $\epsilon\left(t\right)$ appear as functions of time, as the swap probability changes as the protocol evolves. The probability changes only as a function of an external parameter, the splitting  (as opposed to e.g.,\ the current state). Hence Crooks' Theorem is still applicable to thermalizations of this type.

\begin{figure}
\centering
\includegraphics[width=0.5\linewidth]{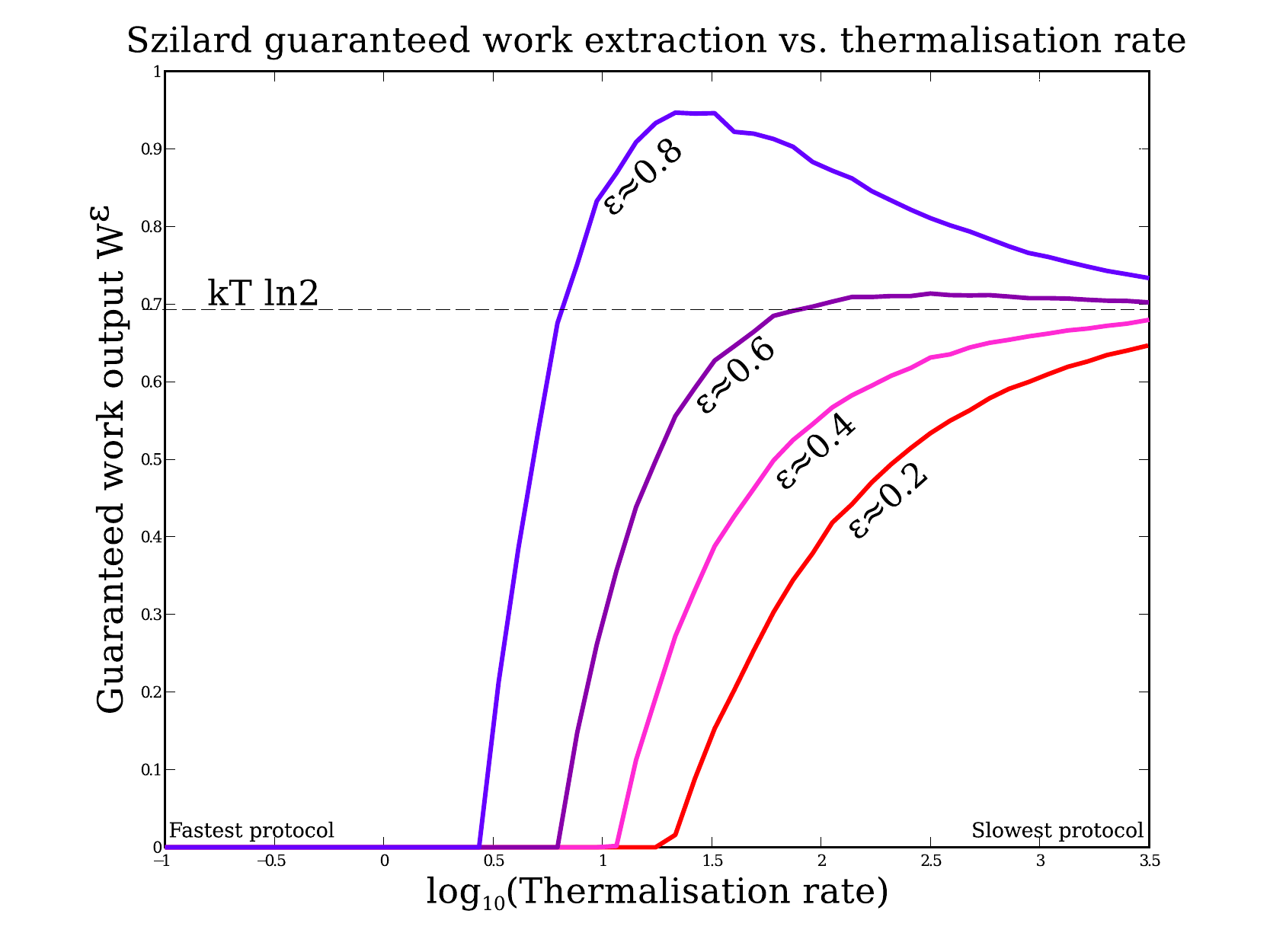}
\caption{Work guaranteed to be extracted from a Szil\'{a}rd engine up to probability $\epsilon$: $w^{\epsilon}$. A Monte Carlo simulation was used to predict the work from the single-electron--box.  $w^{\epsilon}$
approaches $kT\ln2$ as a function of the protocol's speed. For smaller $\epsilon$, $w^{\epsilon}$ approaches from below; and for higher, from above. }
\label{fig:SlicePlot}
\end{figure}

Figure~\ref{fig:SlicePlot} depicts our Monte Carlo simulation.
We randomly generate trajectories by picking a random initial microstate according to the initial-state probability distribution. Then, we evolve the system by small steps, testing at each step if a swap should occur (with probability $p_\mathrm{sw}$). If a swap occurs, we replace the state with a new microstate randomly chosen according to the Gibbs state associated with the current Hamiltonian.
By recording which microstate is occupied when the energy level is raised, we calculate the work cost associated with a particular trajectory.
Repeated runs of the simulation allow us to build up a work distribution, 
to which the results in this paper apply.

\subsection{Analytic expression for the work distribution}
\label{GuaranteedW::sec:1}

The work distribution function for an electron box can also be obtained explicitly from the master equation.

Consider an arbitrary work protocol 
that runs from $t=0$ to $t=\tau$.
The gap is tuned as a function $\epsilon(t)$.
The trajectory $\sigma(t)\in\{0,1\}$ of the system is piece-wise constant,
jumping discontinuously from one energy level to another at some random instants $t_j$ ($j=1,2,   \ldots  $). Therefore, the trajectory is specified uniquely by the initial condition $\sigma_0$, the number $J$ of jumps, and the corresponding instants $t_j$ ($j=1,2,  \ldots ,J$).
The probability distribution function for the trajectory is
\begin{equation}
\label{WorkDist::eq:3}
P_J(t_1,  \ldots ,t_J;\sigma_0)
= \prod_{j=1}^J\Gamma((-1)^{\sigma_0+j+1}\epsilon(t_j))
\exp\left[-S_J(t_1,  \ldots ,t_J;\sigma_0)\right]
\end{equation}
where the \emph{effective action} associated with a given trajectory
has been defined as
\begin{equation}
\label{WorkDist::eq:7}
S_J(t_1,  \ldots ,t_J;\sigma_0) =
\sum_{j=1}^{J+1}\int_{t_{j-1}}^{t_j}{ds}\,
\Gamma((-1)^{\sigma_0+j+1}\epsilon(s))
\end{equation}
and $t_0=0$ and $t_{J+1}=\tau$ are implied.
Checking the normalization is straightforward:
\begin{equation}
\label{WorkDist::eq:8}
P_0(\sigma_0) + \sum_{J=1}^\infty
\prod_{j=1}^J\int_{t_{j-1}}^\tau{dt_j}\,
P_J(t_1,  \ldots ,t_J;\sigma_0) = 1 \,,
\end{equation}
wherein $t_0=0$ is again implied.

The work is  done only while the system is in the state $\sigma=1$.
Hence the contribution to the work along the trajectory is 
\begin{equation}
\label{WorkDist::eq:1}
W_J(t_1,  \ldots,   t_J;\sigma_0)
= \sum_{j=1}^J(-1)^{\sigma_0+j}\epsilon(t_j)
+ (\sigma_0+J\mmod{2})\epsilon_f
- \sigma_0\epsilon_0
\end{equation}
The work distribution function along a trajectory with $J$ jumps is
\begin{equation}
\label{WorkDist::eq:14}
P_J(W;\sigma_0)
= \prod_{j=1}^J\int_{t_{j-1}}^{\tau}{dt_j}\,
P_J(t_1,  \ldots,t_J;\sigma_0)\,
\delta(W-W_J(t_1,  \ldots,t_J;\sigma_0)).
\end{equation}
The total work distribution function can be written in a series
\begin{equation}
\label{WorkDist::eq:16}
P(W)
= p_0e^{-S_0(0)}\delta(W)
+ p_1e^{-S_0(1)}\delta(W-W_c)
+ \sum_{J=1}^\infty\sum_{\sigma_0} p_{\sigma_0}P_J(W;\sigma_0) .
\end{equation}
$P_J(W)$ has a factor of $(\Gamma_0^2e^{-\beta\epsilon})^J$.
At low temperatures, $P_J$ is rapidly suppressed as $J$ increases.

The expression~(\ref{WorkDist::eq:16}) for the work distribution is essentially a perturbative expansion in $\Gamma_0^2$ and converges very quickly for small $\Gamma_0$. For large $\Gamma_0$, however, it becomes impractical to use it for actual calculation because of its slow convergence. Therefore, it will be useful to devise a more general method. We examine the characteristic function $Z(\xi)=\avg{e^{\xi W}}$ of the work distribution function $P(W)$.
We first consider the characteristic function
 $Z_\sigma(\xi)=\avg{e^{\xi W}}_{\sigma}$ conditioned on all trajectories' starting from a definite initial state $\sigma_0$.
Regarded as a function of the operation time $\tau$, $Z_\sigma(\xi;\tau)$ satisfies the master equation~\cite{Imparato05a,Verley13a}
\begin{equation}
\label{GuaranteedW::eq:1}
\partial_\tau Z_\sigma(\lambda;\tau)
= \sum_\sigma\left[\Gamma_{\sigma\sigma'}(\tau)
+ \lambda\partial_\tau\epsilon_\sigma(\tau)\delta_{\sigma\sigma'}\right]
Z_{\sigma'}(\lambda;\tau)
\end{equation}
and the initial condition
\begin{equation}
Z_{  \sigma  }(\xi;0) = e^{\xi\epsilon_\sigma(0)} \,.
\end{equation}
Compared with the original master equation~(\ref{WorkDist::eq:4}) 
for the level-occupation probabilities, the new master equation~(\ref{GuaranteedW::eq:1}) for the characteristic function contains additional diagonal terms.
The full characteristic function is 
\begin{equation}
Z(\xi) = \sum_{\sigma_0}p_{\sigma_0}Z_{\sigma_0}(\xi).
\end{equation}
Recall that $Z(\xi)$ contains the same information as $P(W)$. From $Z(\xi)$, one can calculate $P(W)$ itself and, as shown in Section~\ref{GuaranteedW::sec:2} below, a bound for $D_\infty(P_\fwd(W)\|P_\rev(-W))$.

Let us show that the work distribution in Eq.~(\ref{WorkDist::eq:16}) satisfies  Crooks' fluctuation theorem,
\begin{equation}
\frac{P_\fwd(W)}{P_\rev(-W)} =
\frac{Z_f}{Z_0}e^{\beta W},
\end{equation}
wherein $Z_0$ and $Z_f$ are the partition functions for the initial and final Hamiltonians in the forward protocol. Given a \emph{forward ramping}
$\epsilon(t)$, the \emph{reverse ramping} $\epsilon^\rev(t)$ is defined by
\begin{equation}
\epsilon^\rev(t) = \epsilon(\tau-t) \,.
\end{equation}
In the forward protocol, consider a trajectory $\sigma(t)$ characterized by the initial condition $\sigma_0$, the number $J$ of energy-level jumps and the jump instants $t_j$ ($j=1,2,  \ldots,J$). One can find a unique trajectory $\sigma^\rev(t)$ in the reverse protocol, which is defined by the initial condition
\begin{equation}
\sigma_0^\rev = \sigma_0+J\pmod{2}
\end{equation} and the flip instants
\begin{equation}
t_j^\rev = \tau - t_{J-j+1} \,.
\end{equation}
Note that
\begin{equation}
\epsilon^\rev(t_j^\rev)=\epsilon(t_{J-j+1}).
\end{equation}
The effective action along the reverse trajectory is the same as that along the forward trajectory [cf.~(\ref{WorkDist::eq:7})]:
\begin{equation}
S_J^\rev(t_1^\rev,  \ldots,t_J^\rev;\sigma_0^\rev)
= S_J(t_1,  \ldots,t_J;\sigma_0).
\end{equation}
Further, the work contribution along the reverse trajectory is  the negative of that along the forward trajectory [cf.~(\ref{WorkDist::eq:1})]:
\begin{equation}
W_J^\rev(t_1^\rev,  \ldots,t_J^\rev;\sigma_0^\rev)
= -W_J(t_1,  \ldots,t_J;\sigma_0) \,.
\end{equation}
These observations lead to
\begin{equation}
P_J^\rev(t_1^\rev,  \ldots,t_J^\rev;\sigma_0^\rev)
= P_J(t_1,  \ldots,t_J;\sigma_0)e^{-\beta W_J(t_1,  \ldots,t_J;\sigma_0)}
\exp\left[(\sigma_0+J\mmod{2})\beta\epsilon_f
- \sigma_0\beta\epsilon_0\right]
\end{equation}
and
\begin{equation}
P_J^\rev(-W;\sigma_0^\rev)
= P_J(W;\sigma_0)
e^{-\beta W}
e^{(\sigma_0+J\mmod{2})\beta\epsilon_f
  -\sigma_0\beta\epsilon_0}
\end{equation}
It is then straightforward to prove Crooks' Theorem:
\begin{align*}
  P_\rev(-W)
  & = \frac{1}{1+e^{-\beta\epsilon_0^\rev}}\sum_{J=0}^\infty\sum_{\sigma_0^\rev}
    e^{-\beta\sigma_0^\rev\epsilon_f^\rev}P_J^\rev(-W;\sigma_0^\rev) \\
  & = \frac{1}{1+e^{-\beta\epsilon_f}}\sum_{J=0}^\infty\sum_{\sigma_0}
    e^{-\beta(\sigma_0+J\mmod{2})\epsilon_0} \\
  &{}\qquad\qquad\times
    e^{-\beta W}e^{\beta(\sigma_0+J\mmod{2})\epsilon_f-\sigma_0\beta\epsilon_0}
    P_J(W;\sigma_0) \\
  & = \frac{e^{-\beta W}}{1+e^{-\beta\epsilon_f}}
    \sum_{J=0}^\infty\sum_{\sigma_0}
    e^{-\sigma_0\beta\epsilon_0}P_J(W;\sigma_0) \\
  &= e^{-\beta W}\frac{1+e^{-\beta\epsilon_0}}{1+e^{-\beta\epsilon_f}}
    P_\fwd(W).
\end{align*}

\begin{figure}
\centering
\includegraphics[width=.3\linewidth]{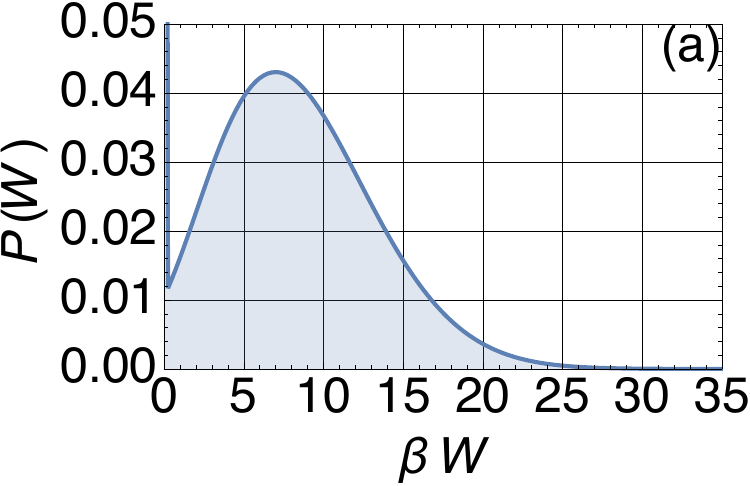}
\includegraphics[width=.3\linewidth]{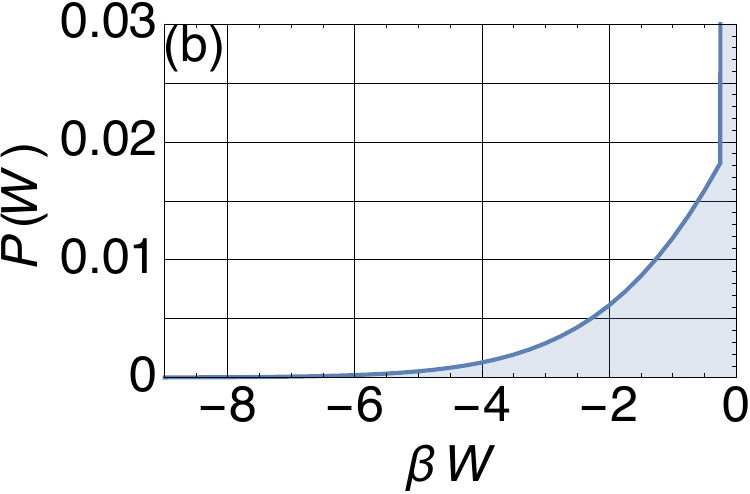}
\caption{Work distributions calculated analytically for the forward (a) and reverse (b) processes on an electron box. 
The two levels initially have the same energy. One level is lifted linearly to $50k_BT$ and then returned to $0$. The values of the zero-energy tunneling rate $\Gamma_0$ and the operation time $\tau$ are set such that $\Gamma_0\tau/\varepsilon_c=k_BT$, wherein $\varepsilon_c$ is the relaxation time of the metallic electrode (charge reservoir).}
\label{GuaranteedW::fig:2}
\end{figure}

 For illustration, examples of the forwards and reverse distributions appear in Fig.~\ref{GuaranteedW::fig:2}. 

\subsection{Upper bound on $D_\infty$}
\label{GuaranteedW::sec:2}

Recall the Markov inequality for a nonnegative random variable $X$:
$$p(X\geq a)\leq \langle X \rangle/a.$$
This is derived by noting that there cannot be too much probability of having a value much greater than the average, or else the average would have to be greater.
In our case, it reads
$$p(w\geq \widetilde{w}^\epsilon) =: \epsilon \leq \langle w \rangle/\widetilde{w}^\epsilon.$$
Thus,
\begin{align}
   \label{eq:MarkovApp}
   \widetilde{w}^\epsilon \leq \langle w \rangle/\epsilon.
\end{align}

\begin{figure}
\centering
\includegraphics[width=0.45\linewidth]{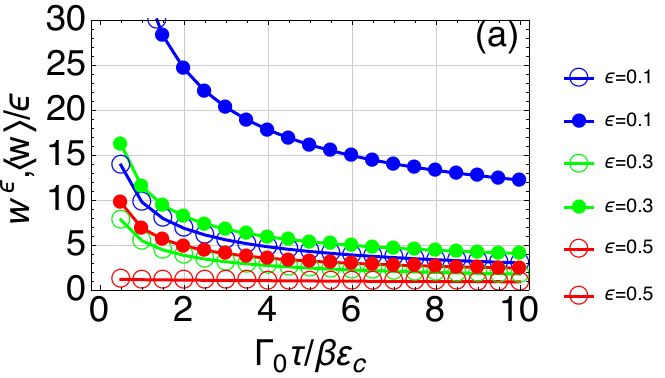}
\includegraphics[width=0.45\linewidth]{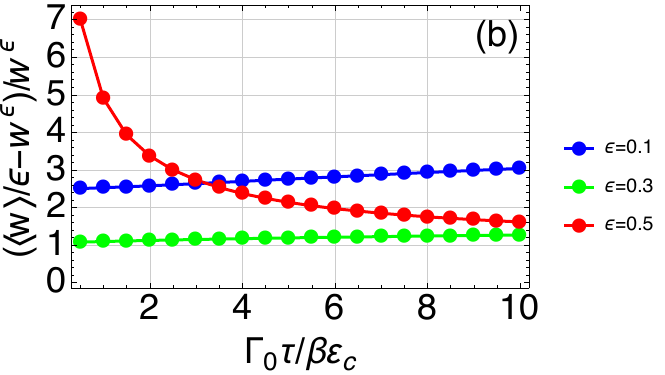}
\caption{$\tilde{w}^\epsilon$ and its upper bounds $\avg{w}/\epsilon$ (which in turn bound $D_\infty$) for different values of $\epsilon$. (a) Individual plots of $\tilde{w}^\epsilon$ and $\avg{w}/\epsilon$. (b) The relative tightness.}
\label{GuaranteedW::fig:1}
\end{figure}

We rearrange the main result, Eq.~\eqref{eq:MainResult}:
\begin{align}  \label{eq:MainResult2b}
D_{\infty} (\widetilde{p}_{\fwd}^{\epsilon}(w) \!|| p_{\rev}(-w))
\!=\!  \beta \widetilde{w}^{\epsilon} 
-\log(\!1\!-\!\epsilon)
+\log Z/\widetilde{Z}.
\end{align}
Substituting in from Ineq.~\eqref{eq:MarkovApp} yields
\begin{align}  \label{eq:MainResult3}
D_{\infty} (\widetilde{p}_{\fwd}^{\epsilon}(w) \!|| p_{\rev}(-w))
\!\leq \!  \beta \langle w \rangle/\epsilon 
-\log(\!1\!-\!\epsilon)
+\log Z/\widetilde{Z} .
\end{align}
(here $\log Z/\widetilde{Z}=\log 2$).
One has only to upper-bound $\avg{w}$. 
$\avg{w}$ can be upper-bounded most easily with the characteristic function $\avg{e^{\lambda w}}$, which bounds $\avg{w}$ due to convexity. This has been illustrated in Fig.~\ref{GuaranteedW::fig:1}.

We finally remark that, as shown in \cite{Aberg13, EgloffDRV12}, in the isothermal limit, the penalty (meaning again the LHS of Eq.~\ref{eq:MainResult2b}), goes to zero. The isothermal limit means that the hopping probabilities multiplying together to give a trajectory's probability as in Eq.~\ref{eq:trajprob} take the form of thermal state occupation probabilities $\gamma_j^{i_j}:=\exp{-\beta E(\ket{i_j,\lambda_j})}/Z(\lambda_j)$. The probability of a forwards trajectory becomes $p(traj)=\gamma_0^{i_0}\gamma_1^{i_1}...\gamma_f^{i_f}$ whereas the reverse trajectory has the probability $p(traj-inv)=\gamma_f^{i_f}\gamma_f^{i_{f-1}}...\gamma_1^{i_0}$. The probability of a given time sequence of work values from the elementary steps is then a product of individual distributions $p(w_1,w_2,...)=p_1(w_1)p_2(w_2)... $. This allows one to use the McDiarmid inequality for independent random variables as in~\cite{EgloffDRV12} to show that there is concentration around the average in the limit of breaking up an isothermal time evolution into infinitely many substeps of energy shifts. If we write $\widetilde{w}^{\epsilon}= \langle w \rangle+\varepsilon$, both $\epsilon$ and $\varepsilon$ tend towards zero in this limit. Combining that with the also known fact that  $\langle w \rangle =-kT\log Z/\widetilde{Z}$ in the isothermal case, we see that the RHS of Eq.~\ref{eq:MainResult2b} tends to zero in this limit; thus the LHS also tends to zero. 
To find, for a given initial state and initial and final Hamiltonians, a protocol such that this penalty tends to zero, one can accordingly begin the protocol with lifting the OUT levels to the corresponding thermal levels, and then perform isothermal quasistatic extraction as described above. The lifting of the OUT levels then undoes their initial lowering in the $\sim$-protocol, undoing any work cost and returning the state to being a thermal state.  This limit also illustrates why the $\epsilon$-versions of the worst-case work and associated penalty are physically natural to introduce. Strictly speaking the worst case work $w^0$ could be much larger than the average even in the isothermal case, but the probability of this happening can be arbitrarily small.

\end{document}